\documentclass[twocolumn,superscriptaddress,
showpacs,preprintnumbers,amsmath,amssymb]{revtex4}

\usepackage{amssymb,amsmath,amsthm,amsfonts,amstext,amsbsy,amscd}
\usepackage{graphicx,epsfig} %Include figure files
\usepackage{mathrsfs}
\usepackage{multirow}
\usepackage{color} %text color, etc.

\newtheorem{Thm}{Theorem}%[section]
\newtheorem{Lem}[Thm]{Lemma}
\newtheorem{Prop}[Thm]{Proposition}
\newtheorem{Cor}[Thm]{Corollary}
\theoremstyle{definition}

\newtheorem{Rem}[Thm]{Remark}

\newcommand{\ket}[1]{\left|{#1}\right\rangle}
\newcommand{\bra}[1]{\left\langle{#1}\right|}
\newcommand{\Tr}{\mathop{\mathrm{Tr}}\nolimits} 
\newcommand{\inn}[2]{\left\langle #1 | #2 \right\rangle}

\def\Complex{\mathbb{C}}
\newcommand*{\cC}{\mathcal{C}}
\newcommand*{\cE}{\mathcal{E}}

\newcommand*{\cI}{\mathcal{I}}

\newcommand*{\cL}{\mathcal{L}}

\newcommand*{\cN}{\mathcal{N}}

\begin{document}

%%%%%%      title      %%%%%%
\title{Activation of zero-error classical capacity in low-dimensional quantum systems}

\author{Jeonghoon Park}\email{zucht@korea.ac.kr}
\affiliation{
Smart Quantum Communication Research Center, Korea University, Seoul 02841, Korea
}
\author{Jun Heo}\email{junheo@korea.ac.kr}
\affiliation{
School of Electrical Engineering, Korea University, Seoul 02841, Korea
}

\date{\today}

%%%%%%%%%%%%%%%%%%%%%%%%%%%%%%%%%%%%%%%%%%%%
%%                                                            Abstract                                                    %%
%%%%%%%%%%%%%%%%%%%%%%%%%%%%%%%%%%%%%%%%%%%%
\begin{abstract}
Channel capacities of quantum channels can be nonadditive 
even if one of two quantum channels has no channel capacity. 
We call this phenomenon \emph{activation} of the channel capacity. 
In this paper, we show that 
when we use a quantum channel on a qubit system, 
only a noiseless qubit channel can generate 
the activation of the zero-error classical capacity. 
In particular, we show that 
the zero-error classical capacity of two quantum channels on qubit systems 
cannot be activated. 
Furthermore, we present a class of examples 
showing the activation of the zero-error classical capacity 
in low-dimensional systems. 
\end{abstract}


\maketitle

%%%%%%%%%%%%%%%%%%%%%%%%%%%%%%%%%%%%%%%%%%%%
%%                                                      Introduction                                                    %%
%%%%%%%%%%%%%%%%%%%%%%%%%%%%%%%%%%%%%%%%%%%%
\section{Introduction}
% Nonadditivity in classical 
Zero-error channel capacity is quite different with the ordinary channel capacity, 
and it has unique properties in both classical and quantum systems
~\cite{KO98,GAM16}. 
In particular, 
while the ordinary capacity of classical channels is additive, 
the zero-error capacity $\cC_0$ of classical channels is nonadditive
~\cite{Haemers79}: 
there are two classical channels $\cE_1$ and $\cE_2$ such that 
\begin{equation*}
\cC_0(\cE_1\times\cE_2)>\cC_0(\cE_1)+\cC_0(\cE_2) 
~\text{and}~\cC_0(\cE_{1,2})>0.
\end{equation*}

% Activation 
In quantum systems, 
a stronger form of nonadditivity is possible~\cite{Duan09}: 
there is a quantum channel $\cN$ with $\cC_0(\cN)=0$ such that 
\begin{equation*}
\cC_0(\cI_2\otimes\cN)>\cC_0(\cI_2)+\cC_0(\cN), 
\end{equation*}
where $\cI_2$ is a noiseless qubit channel. 
We may think that 
a noiseless qubit channel $\cI_2$ activates 
the ability of the useless channel $\cN$ to transmit classical information, 
and so we call this nonadditivity \emph{activation}. 
However, for classical channels $\cE_{1,2}$, 
the condition that $\cC_0(\cE_1)=0$ implies $\cC_0(\cE_1\times\cE_2)=\cC_0(\cE_2)$. 
Thus, the activation is a quantum phenomenon 
which never occur in classical channels. 

% Superactivation
Furthermore, nonadditivity can happen 
even when two channels have no capacities, 
and such a nonadditivity is called \emph{superactivation}~\cite{Duan09,CCH11}. 
However, the superactivation of the zero-error classical capacity 
can occur only in high-dimensional quantum systems; 
indeed, the input dimensions of quantum channels must be 
greater than or equal to 4~\cite{PL12,SS15}.
% Motivation 
It has been known that 
many extraordinary features of quantum channel capacities are revealed 
in high-dimensional 
or sufficiently large dimensional quantum systems~\cite{Hastings09,SY08,SS09a}, 
but it is not clear that those features can also happen in low-dimensional cases.
Hence, it could be important to concern 
how quantum phenomena occur in low-dimensional quantum systems, 
especially qubit systems. 

% Results & meanings %
We here take into account 
the zero-error classical capacity of quantum channels 
and investigate the activation of the zero-error classical capacity 
in low-dimensional quantum systems. 
We show that 
when we have a quantum channel $\cN$ on a qubit system, 
$\cN$ must be noiseless to be activated; 
that is, 
only a noiseless qubit channel can cause the activation. 
In addition, we show that 
the zero-error classical capacity of quantum channels cannot be activated 
when the quantum channels are on qubit systems. 
Moreover, we construct a class of examples 
showing the activation of the zero-error classical capacity. 
In particular, 
we present an example of activation 
which has the smallest input dimensions so far. 

% organized % 
This paper is organized as follows. 
In Sec.~\ref{sec:Act_(2,n)}, 
we present when the activation happens 
if the input dimension of one channel is 2. 
In Sec.~\ref{sec:Act_(2,2)}, 
we show that 
two quantum channels on qubit systems cannot generate the activation. 
In Sec.~\ref{sec:Act_(2,4)}, 
we construct examples which show the activation of the zero-error classical capacity
in low-dimensional input systems. 
Finally, we summarize our results in Sec.~\ref{sec:conclusion}.

%%%%%%%%%%%%%%%%%%%%%%%%%%%%%%%%%%%%%%%%%%%%
%%                                                     Act. in 2*n                                              %%
%%%%%%%%%%%%%%%%%%%%%%%%%%%%%%%%%%%%%%%%%%%%
\section{Necessary conditions for activation}\label{sec:Act_(2,n)}
In this section, 
we investigate the activation of the zero-error classical capacity 
when we have a quantum channel on $\Complex^2$. 
We show that 
the activation cannot happen 
unless the quantum channel on $\Complex^2$ is noiseless. 

% def of ZEC / noncommutative graph %
For a quantum channel $\cN$ with Kraus operators $E_{i}$, 
the one-shot zero-error classical capacity $\cC_0^{(1)}(\cN)$ is defined as 
\begin{equation*}
\cC_0^{(1)}(\cN)\equiv\log{\alpha(\cN)},
\end{equation*}
where $\alpha(\cN)$ is the maximum number of (orthogonal) vectors 
$\ket{\psi_{1}}, \dots, \ket{\psi_{m}}$ 
such that 
\begin{equation}\label{eq:codeword}
\ket{\psi_{s}}\bra{\psi_{t}} \perp S 
\equiv\mathrm{span}\{E_{i}^{\dag}E_{j}: i,j\}, \forall{s}\ne{t}. 
\end{equation}
The (asymptotic) zero-error classical capacity $\cC_0(\cN)$ is defined by 
\begin{equation}\label{eq:C_0}
\cC_0(\cN)
\equiv\lim_{n\rightarrow\infty}\frac{\cC_0^{(1)}(\cN^{\otimes{n}})}{n}
=\lim_{n\rightarrow\infty}\frac{\cC_0^{(1)}(S^{\otimes{n}})}{n}.
\end{equation}
We note that 
$\cC_0^{(1)}(\cN)$ depends only on its associated subspace $S$ 
called the \emph{noncommutative graph} of $\cN$~\cite{DSW13}. 

\begin{Rem}\label{Rem:noncomm_graph}
The noncommutative graph $S(\cN)$ of a quantum channel $\cN$ has two properties: 
$S(\cN)=S(\cN)^{\dag}$ and $I\in{S(\cN)}$. 
Conversely, 
any subspace $S\le\cL(\Complex^{n})$ such that $S=S^{\dag}$ and $I\in{S}$ is indeed 
a noncommutative graph of some quantum channel~\cite{DSW13}. 
Hence, we will use $\cC_0^{(1)}(S)$ and $\cC_0(S)$ 
without referring to any specific quantum channel. 
\end{Rem}

\begin{Rem}\label{Rem:bit_unit}
In order to measure in the number of bits, 
we define $\cC_0^{(1)}(\cN)$ as $\log{\alpha(\cN)}$, 
where the base of the logarithm is 2. 
\end{Rem}

We can easily obtain from Eq.~(\ref{eq:codeword}) 
the following characterization~\cite{Duan09}.  
\begin{Prop}\label{Prop:NoRankOne}
Let $S$ be a noncommutative graph. 
Then $\cC_0^{(1)}(S)=0$ 
if and only if $S^{\perp}$ has no rank-one matrices. 
\end{Prop}

% def. of Act. %
Let $S$ and $T$ be noncommutative graphs, 
and $\cC_0^{(1)}(T)=0$. 
The one-shot zero-error classical capacity of $S$ and $T$ can be \emph{activated} 
if and only if 
\[
\cC_0^{(1)}(S\otimes{T})>\cC_0^{(1)}(S)+\cC_0^{(1)}(T). 
\]

We can see the noncommutative graph $\cL(\Complex^n)$ 
as an extremely noisy channel, 
and indeed $\cC_0^{(1)}(\cL(\Complex^n))=0$ 
by Proposition~\ref{Prop:NoRankOne}. 
However, the following theorem says that 
$\cL(\Complex^n)$ cannot cause the activation. 
Thus, we may think that 
channels should not be too noisy 
in order to be activated. 
\begin{Thm}\label{Thm:CompletelyNoisy}
For any noncommutative graph $S$, 
$\cC_0^{(1)}(S\otimes\cL(\Complex^n))=\cC_0^{(1)}(S)+\cC_0^{(1)}(\cL(\Complex^n))$.
\end{Thm}
\begin{proof}
If $\cC_0^{(1)}(S\otimes\cL(\Complex^n))=0$, 
clearly, $\cC_0^{(1)}(S)=0$. 

We assume that $\cC_0^{(1)}(S\otimes\cL(\Complex^n))>0$. 
For any $\ket{\Phi}$ and $\ket{\Psi}$ satisfying 
\[
\ket{\Phi}\bra{\Psi} \perp S\otimes\cL(\Complex^n),
\]
let 
\begin{eqnarray*}
\ket{\Phi} &=& \sum_{i}\sqrt{\lambda_i}\ket{\lambda_i}\ket{\phi_i} \\
\ket{\Psi} &=& \sum_{j}\sqrt{\mu_j}\ket{\mu_j}\ket{\psi_j}
\end{eqnarray*}
in the Schmidt decompositions, 
where $\lambda_i$\rq{}s and $\mu_j$\rq{}s are positive. 
Then for any $A\in{S}$ and $s$, $t$, 
\begin{eqnarray*}
0 &=& \Tr[(\ket{\Phi}\bra{\Psi})^{\dag}(A\otimes\ket{\phi_s}\bra{\psi_t})] \\
&=& \sum_{i,j}\sqrt{\lambda_{i}\mu_{j}}\bra{\lambda_i}A\ket{\mu_j}\inn{\phi_i}{\phi_s}\inn{\psi_t}{\psi_j} \\
&=& \sqrt{\lambda_{s}\mu_{t}}\bra{\lambda_s}A\ket{\mu_t}.
\end{eqnarray*}
So, $\ket{\lambda_s}\bra{\mu_t}\perp{S}$ for any $s, t$. 
Thus, $\cC_0^{(1)}(S\otimes\cL(\Complex^n))\le\cC_0^{(1)}(S)$, 
and hence $\cC_0^{(1)}(S\otimes\cL(\Complex^n))=\cC_0^{(1)}(S)$. 
\end{proof}

Now, we consider noncommutative graphs $S$ and $T$ 
in $\cL(\Complex^2)$ and $\cL(\Complex^n)$, respectively. 
Using Proposition~\ref{Prop:NoRankOne}, 
we can see the following proposition 
which is used in the proof of Theorem~\ref{Thm:Act_2*n}. 
\begin{Prop}\label{Prop:Act_basic}
Let $S\le\cL(\Complex^2)$ and $T\le\cL(\Complex^n)$ be noncommutative graphs 
and $\cC_0^{(1)}(T)=0$. 
Suppose that 
$\ket{\psi_i}=\ket{0}\ket{v_i}+\ket{1}\ket{w_i}$ are orthogonal states 
satisfying Eq.~(\ref{eq:codeword}) 
with respect to $S\otimes{T}$, 
where $\ket{v_i}, \ket{w_i} \in\Complex^n$ and $1\le{i}\le3$. 
Then for any ${i}\ne{j}$, 
\begin{enumerate}
\item[(i)] $\ket{v_i}$'s and $\ket{w_i}$'s are nonzero. 
\item[(ii)] $\ket{v_i}$ and $\ket{w_i}$ are linearly independent. 
\item[(iii)] $\ket{w_i}$ and $\ket{w_j}$ are linearly independent. 
\end{enumerate}
\end{Prop}

\begin{proof}
Since $I_2\in{S}$ and 
$\ket{\psi_i}=\ket{0}\ket{v_i}+\ket{1}\ket{w_i}$ satisfy Eq.~(\ref{eq:codeword}) 
with respect to $S\otimes{T}$, %아니면 From the assumption, ? 
\[
A_{ij}\equiv\ket{v_i}\bra{v_j}+\ket{w_i}\bra{w_j}\in{T^{\perp}} 
\]
for any $1\le{i}\ne{j}\le3$. 
We note that $T^{\perp}$ has no rank-one matrices 
from Proposition~\ref{Prop:NoRankOne}. 

(i) Suppose that $\ket{v_i}=0$, say $i=1$. 
Then $A_{1j}=\ket{w_1}\bra{w_j}\in{T^{\perp}}$ 
for $j=2,3$. 
Since $\ket{\psi_1}\ne0$, 
$\ket{w_1}\ne0$, 
and so $\ket{w_2}=0=\ket{w_3}$. 
Then $A_{23}=\ket{v_2}\bra{v_3}\in{T^{\perp}}$, 
and hence $\ket{v_2}=0$ or $\ket{v_3}=0$. 
Thus, $\ket{\psi_2}=0$ or $\ket{\psi_3}=0$ 
which is a contradiction. 
Similarly, we can see that $\ket{w_i}$\rq{}s are nonzero. 

(ii) Suppose that $\ket{v_1}=\alpha\ket{w_1}$ 
for some $\alpha$. 
Then $A_{1j}=\ket{w_1}(\alpha\bra{v_j}+\bra{w_j})\in{T^{\perp}}$ 
for $j=2,3$, 
and so $\alpha\bra{v_j}+\bra{w_j}=0$ by (i). 
Then $A_{23}=(1+|\alpha|^2)\ket{v_2}\bra{v_3}\in{T^{\perp}}$, 
and hence $\ket{v_2}=0$ or $\ket{v_3}=0$ 
which is a contradiction to (i). 

(iii) Suppose that $a\ket{w_i}+b\ket{w_j}=0$. 
Then $aA_{ik}+bA_{jk}=(a\ket{v_i}+b\ket{v_j})\bra{v_k}\in{T^{\perp}}$. 
By (i), 
$a\ket{v_i}+b\ket{v_j}=0$, 
and so $a\ket{\psi_i}+b\ket{\psi_j}=0$. 
Since $\ket{\psi_i}$ and $\ket{\psi_j}$ are linearly independent, 
$a=0=b$, 
and hence $\ket{w_i}$ and $\ket{w_j}$ are linearly independent. 
\end{proof}

We note that $\Complex{I_2}$ is associated with 
a noiseless qubit channel (up to unitary equivalence). 
The following theorem says that 
the only quantum channel on $\Complex^2$ causing the activation is 
a noiseless qubit channel. 
\begin{Thm}\label{Thm:Act_2*n}
Let $S$ and $T$ be noncommutative graphs 
in $\cL(\Complex^2)$ and $\cL(\Complex^n)$, respectively. 
If the one-shot zero-error classical capacity of $S$ and $T$ can be activated, 
then $S={\Complex}I_2$. 
\end{Thm}
\begin{proof}
Since any qubit channel cannot cause 
the superactivation of the zero-error classical capacity
~\cite{PL12}, 
$\cC_0^{(1)}(S)>0$ or $\cC_0^{(1)}(T)>0$. 
Moreover, we note that any noncommutative graph in $\cL(\Complex^2)$ is 
${\Complex}I_2$, $\mathrm{span}\{I_2, \sigma_3\}$, 
$\mathrm{span}\{I_2, \sigma_1, \sigma_3\}$, or $\cL(\Complex^2)$ 
(up to unitary equivalence), 
and its one-shot zero-error classical capacity is 
$1$, $1$, $0$, $0$, respectively~\cite{DSW13}. 

We first assume that $\cC_0^{(1)}(S)=0<\cC_0^{(1)}(T)$. 
When $S=\cL(\Complex^2)$, 
by Theorem~\ref{Thm:CompletelyNoisy}, 
$\cC_0^{(1)}(S\otimes{T})=\cC_0^{(1)}(S)+\cC_0^{(1)}(T)$. 
Suppose that $S=\mathrm{span}\{I_2, \sigma_1, \sigma_3\}$. 
Let 
\[
\ket{\psi_i}=\ket{0}\ket{v_i}+\ket{1}\ket{w_i} \in\Complex^2\otimes\Complex^n
\] 
satisfy Eq.~(\ref{eq:codeword}) 
with respect to $S\otimes{T}$. 
Then for any $i\ne{j}$, 
\[
\bra{\psi_i}(P\otimes{Q})\ket{\psi_j}=0, \forall P\in{S}, Q\in{T}.
\]
Choosing $P=I_2, \sigma_3$, and $\sigma_1$, 
we can see that 
\begin{eqnarray}
\ket{v_i}\bra{v_j}\pm\ket{w_i}\bra{w_j} &\perp& T, \label{eq:perp1} \\
\ket{v_i}\bra{w_j}+\ket{w_i}\bra{v_j} &\perp& T. \label{eq:perp2}
\end{eqnarray}
Define 
\begin{equation*}
\ket{\phi_i}=\left\{
\begin{array}{ll}
\ket{v_i} & \text{if}~\ket{v_i}\ne0\\
\ket{w_i} & \text{if}~\ket{v_i}=0.
\end{array} \right.
\end{equation*}
Then from Eqs.~(\ref{eq:perp1})~and~(\ref{eq:perp2}), 
we can see that $\ket{\phi_i}\bra{\phi_j}\perp{T}$ 
for any $i\ne{j}$. 
Thus, 
$\cC_0^{(1)}(S\otimes{T})\le\cC_0^{(1)}(T)$, 
and hence $\cC_0^{(1)}(S\otimes{T})=\cC_0^{(1)}(T)$. 

We now assume that $\cC_0^{(1)}(S)>0=\cC_0^{(1)}(T)$. 
Suppose that $S=\mathrm{span}\{I_2, \sigma_3\}$. 
Then there are 
\[
\ket{\psi_i}=\ket{0}\ket{v_i}+\ket{1}\ket{w_i} \in\Complex^2\otimes\Complex^n
\]
satisfying Eq.~(\ref{eq:codeword}) 
with respect to $S\otimes{T}$, 
where $1\le{i}\le3$. 
It is not hard to show that 
for any $i\ne{j}$, 
\[
\ket{v_i}\bra{v_j}\pm\ket{w_i}\bra{w_j}\in{T^\perp}, 
\]
and so $\ket{v_i}\bra{v_j}\in{T^\perp}$. 
Since $\cC_0^{(1)}(T)=0$, 
by Proposition~\ref{Prop:NoRankOne}, 
$T^{\perp}$ has no rank-one matrices. 
Thus, $\ket{v_i}=0$ or $\ket{v_j}=0$. 
This is a contradiction by Proposition~\ref{Prop:Act_basic}, 
and we can conclude that $S={\Complex}I_2$.
\end{proof}

A necessary condition of activation in Theorem~\ref{Thm:Act_2*n} 
can be extended to the asymptotic case. 
%The result in Theorem~\ref{Thm:Act_2*n} can be extended to the asymptotic case. 

\begin{Thm}
Let $S$ and $T$ be noncommutative graphs 
in $\cL(\Complex^2)$ and $\cL(\Complex^n)$, respectively. 
If the zero-error classical capacity of $S$ and $T$ can be activated, 
then $S={\Complex}I_2$. 
\end{Thm}

\begin{proof}
We note that 
$\cC_0(S)=\cC_0^{(1)}(S)$
%the (asymptotic) zero-error classical capacity $\cC_0(S)$ equals 
%its one-shot zero-error classical capacity $\cC_0^{(1)}(S)$ 
for any noncommutative graph $S\le\cL(\Complex^2)$~\cite{DSW13}. 
%Thus, $\cC_0^{(1)}(S^{\otimes{k}})=k\cdot\cC_0^{(1)}(S)$ 
%for any $k\ge1$. 
%
For the case of $\cC_0(S)=0<\cC_0(T)$, 
applying Theorem~\ref{Thm:Act_2*n} recursively, 
we obtain 
\[
\cC_0^{(1)}(S^{\otimes{k}}\otimes{T^{\otimes{k}}})
=\cC_0^{(1)}(S^{\otimes{(k-1)}}\otimes{T^{\otimes{k}}})
=\cdots=\cC_0^{(1)}(T^{\otimes{k}}).
\]
Hence, $\cC_0(S\otimes{T})=\cC_0(T)$. 

We now consider the case of $\cC_0(S)>0=\cC_0(T)$, 
where $S=\mathrm{span}\{I_2, \sigma_3\}$. 
We will show that 
\begin{equation}\label{eq:claim}
\cC_0^{(1)}(S^{\otimes{k}}\otimes{T})=\cC_0^{(1)}(S^{\otimes{k}})
\end{equation}
which implies $\cC_0(S\otimes{T})=\cC_0(S)$. 
Let 
\[
\ket{\psi_i}=\sum_{t=0}^{2^k-1}\ket{t}\ket{v_{i,t}}\in\Complex^{2^k}\otimes\Complex^n
\]
satisfy Eq.~(\ref{eq:codeword}) with respect to $S^{\otimes{k}}\otimes{T}$. 
We note that 
\[
S^{\otimes{k}}=\mathrm{span}\{I_2, \sigma_3\}^{\otimes{k}}
=\mathrm{span}\{\ket{t}\bra{t}: 0\le{t}\le2^k-1\}. 
\]
Then for any $i\ne{j}$, 
\[
0=\bra{\psi_i}\left(\ket{t}\bra{t}\otimes{M}\right)\ket{\psi_j}
=\bra{v_{i,t}}M\ket{v_{j,t}}
\]
for any $t$ and $M\in{T}$. 
By Proposition~\ref{Prop:NoRankOne}, 
\begin{equation}\label{eq:zeroterms}
\ket{v_{i,t}}\bra{v_{j,t}}=0
\end{equation}
for any $t$ and $i\ne{j}$. 

We now use the induction on $k$ to prove Eq.~(\ref{eq:claim}). 
For $k=1$, 
it holds by Theorem~\ref{Thm:Act_2*n}. 
Assume that Eq.~(\ref{eq:claim}) holds for $k\ge1$.
Suppose that 
\[
\cC_0^{(1)}(S^{\otimes{(k+1)}}\otimes{T})>\cC_0^{(1)}(S^{\otimes{(k+1)}})
\]
Then there exist orthogonal vectors 
\[
\ket{\psi_i}=
\sum_{t=0}^{2^{k+1}-1}\ket{t}\ket{v_{i,t}}\in\Complex^{2^{k+1}}\otimes\Complex^n
\]
satisfying Eq.~(\ref{eq:codeword}) with respect to $S^{\otimes{(k+1)}}\otimes{T}$, 
where $i=1,\dots,2^{k+1}+1$. 
Let us consider 
\begin{eqnarray*}
\ket{\psi_i}_u &\equiv& \sum_{d=0}^{2^{k}-1}\ket{d}\ket{v_{i,d}} 
\in\Complex^{2^k}\otimes\Complex^n, \\
\ket{\psi_i}_l &\equiv& \sum_{d=2^{k}}^{2^{k+1}-1}\ket{d-2^{k}}\ket{v_{i,d}} 
\in\Complex^{2^k}\otimes\Complex^n. 
\end{eqnarray*}
Since $\cC_0^{(1)}(S^{\otimes{k}}\otimes{T})=\cC_0^{(1)}(S^{\otimes{k}})=k$, 
by Eq.~(\ref{eq:zeroterms}), 
there exist at least $(2^{k}+1)$ zero $\ket{\psi_i}_u$'s. 
However, 
$\ket{\psi_i}_l$'s, for which $\ket{\psi_i}_u$'s are zero, 
are nonzero and 
satisfy Eq.~(\ref{eq:codeword}) with respect to $S^{\otimes{k}}\otimes{T}$. 
Therefore, 
\[
\cC_0^{(1)}(S^{\otimes{k}}\otimes{T})
\ge\log(2^k+1)>k=\cC_0^{(1)}(S^{\otimes{k}}). 
\]
This is a contradiction to the induction hypothesis, 
and hence Eq.~(\ref{eq:claim}) holds for all $k$. 
\end{proof}

\begin{Rem}
Some noisy qubit channel can have a positive zero-error classical capacity; 
for example, the dephasing channel $\cN(\rho)=(1-p)\rho+p{\sigma_2}\rho{\sigma_2}$, 
where $0<p<1$. 
However, by Theorem~\ref{Thm:Act_2*n}, 
such a noisy qubit channel cannot generate the activation 
even with a small amount of noise. 
\end{Rem}

%%%                                        2*2,                                             %%%
%%%%%%%%%%%%%%%%%%%%%%%%%%%%%%%%%%%%%%%%%
\section{Nonactivation on $\Complex^{2}\otimes\Complex^{2}$}\label{sec:Act_(2,2)}
We here show that 
the one-shot zero-error classical capacity of two quantum channels 
on $\Complex^{2}$ cannot be activated. 

%\begin{Thm}\label{Thm:Act_2*2}
%Let $S\le\cL(\Complex^2)$ be a noncommutative graph with $\cC_0^{(1)}(S)=0$. 
%Then $\cC_0^{(1)}({\Complex}I_2\otimes S)=1=\cC_0^{(1)}({\Complex}I_2)$.
%\end{Thm}
%%
%\begin{proof}
%Assume that $\cC_0^{(1)}({\Complex}I_2\otimes S)\ge\log{3}$. 
%Then there are 
%\[
%\ket{\psi_i}=\ket{0}\ket{v_i}+\ket{1}\ket{w_i} \in\Complex^2\otimes\Complex^2
%\]
%satisfying Eq.~(\ref{eq:codeword}) 
%with respect to ${\Complex}I_2\otimes{S}$, 
%where $1\le{i}\le3$. 
%%
%Define 
%\[
%A_{ij}\equiv\ket{v_i}\bra{v_j}+\ket{w_i}\bra{w_j}.
%\]
%\end{proof}

\begin{Cor}\label{Cor:NonAct_2*2}
For any pair of quantum channels whose input  systems are on $\Complex^{2}$, 
the one-shot zero-error classical capacity cannot be activated. 
%\textcolor{red}{
%that is, 
%for any noncommutative graphs $S$ and $T$ in $\cL(\Complex^2)$ 
%with $\cC_0^{(1)}(S)=0$, 
%$\cC_0^{(1)}(S\otimes{T})=\cC_0^{(1)}(T)$. 
%}
\end{Cor}

\begin{proof}
Since no qubit channel can cause 
the superactivation of the one-shot zero-error classical capacity~\cite{PL12}, 
let $S$ and $T$ be noncommutative graphs in $\cL(\Complex^2)$, 
and let $\cC_0^{(1)}(S)=0<\cC_0^{(1)}(T)$. 
%In order that 
%the one-shot zero-error classical capacity $\cC_0^{(1)}(S\otimes{T})$ can be activated, 
%from Theorem~\ref{Thm:Act_2*n}, 
%$S={\Complex}I_2$, 
In the first part of the proof of Theorem~\ref{Thm:Act_2*n}, 
we have shown that 
if a noncommutative graph $S\le\cL(\Complex^2)$ has $\cC_0^{(1)}(S)=0$, 
then 
\begin{equation}\label{eq:nonAct}
\cC_0^{(1)}(S\otimes{\widetilde{T}})=\cC_0^{(1)}(\widetilde{T})
\end{equation} 
for any noncommutative graph $\widetilde{T}\in\cL(\Complex^n)$ 
with $\cC_0^{(1)}(\widetilde{T})>0$. 
Hence, $\cC_0^{(1)}(S\otimes{T})=\cC_0^{(1)}(T)$. 
\end{proof}

We can see that 
Corollary~\ref{Cor:NonAct_2*2} can be extended to the case of asymptotic capacity. 
Let $S$ and $T$ be noncommutative graphs in $\cL(\Complex^2)$ 
with $\cC_0(S)=0<\cC_0(T)$. 
Then we can show that 
\begin{eqnarray*}
\cC_0^{(1)}\left((S\otimes{T})^{\otimes{k}}\right)
%&=&\cC_0^{(1)}(S^{\otimes{k}}\otimes{T^{\otimes{k}}}) \\
&=&\cC_0^{(1)}\left(S\otimes(S^{\otimes{(k-1)}}\otimes{T^{\otimes{k}}})\right) \\
&=&\cC_0^{(1)}\left(S^{\otimes{(k-1)}}\otimes{T^{\otimes{k}}}\right) \\
&=&\cdots \\
&=&\cC_0^{(1)}\left(T^{\otimes{k}}\right) 
\end{eqnarray*}
by applying Eq.~(\ref{eq:nonAct}) recursively. 
Hence, we can see that 
$\cC_0(S\otimes{T})=\cC_0(T)$. 
Therefore, we obtain the following corollary. 

\begin{Cor}\label{Cor:NonAct_2*2_asymp}
For any pair of quantum channels whose input  systems are on $\Complex^{2}$, 
the zero-error classical capacity cannot be activated. 
%that is, 
%for any noncommutative graphs $S$ and $T$ in $\cL(\Complex^2)$ 
%with $\cC_0(S)=0$, 
%$\cC_0(S\otimes{T})=\cC_0(T)$. 
\end{Cor}

\begin{Rem}
We can view that 
the results in corollaries~\ref{Cor:NonAct_2*2} and~\ref{Cor:NonAct_2*2_asymp} are 
an extension of the results in Ref.~\cite{PL12}, 
in which it was shown that 
any qubit channel cannot cause 
the superactivation of $\cC_0^{(1)}$ and $\cC_0$. 
\end{Rem}

%%%%%%%%%%%%%%%%%%%%%%%%%%%%%%%%%%%%%%%%%%%%
%%                                                       Act eg on 2*4                                                  %%
%%%%%%%%%%%%%%%%%%%%%%%%%%%%%%%%%%%%%%%%%%%%
\section{A class of examples}
\label{sec:Act_(2,4)} 
In this section, 
we construct noncommutative graphs 
which generate the activation of the zero-error classical capacity. 
\begin{Thm}~\label{Thm:Act_eg}
For each $m\ge3$, 
there is a noncommutative graph $T\le\cL(\Complex^{m+1})$ 
with $\cC_0^{(1)}(T)=0$ 
such that $\cC_0^{(1)}(\Complex{I_2}\otimes{T})\ge\log{m}>1=\cC_0^{(1)}(\Complex{I_2})+\cC_0^{(1)}(T)$.
\end{Thm}
\begin{proof}
Define 
\[
B_{ij} \equiv \ket{i}\bra{j}+\ket{i+1}\bra{j+1}\in\Complex^{(m+1)\times(m+1)},
\]
where $0\le{i}\ne{j}\le{m-1}$. 
Then we can easily see that 
\[
T\equiv\mathrm{span}\{B_{ij}: 0\le{i}\ne{j}\le{m-1}\}^{\perp}
\]
is a noncommutative graph. 

We first show that 
\[
T^{\perp}=\mathrm{span}\{B_{ij}: 0\le{i}\ne{j}\le{m-1}\}
\]
has no rank-one matrices. 
{\it i.e.,} $\cC_0^{(1)}(T)=0$. 
Assume to the contrary that 
$T^{\perp}$ has a rank-one matrix 
\begin{equation*}
B\equiv\sum_{0\le{i}\ne{j}\le{m-1}}\alpha_{ij}B_{ij}
\end{equation*}
for some $\alpha_{ij}$. 
On the other hand, 
we can write 
\[
B=\ket{\psi}\bra{\phi}, 
\]
where $\ket{\psi}=\sum_{i=0}^{m}a_i\ket{i}$ and $\ket{\phi}=\sum_{i=0}^{m}b_i^*\ket{i}$ are nonzero vectors. 
Let the $q$th column of $B$ be the right most nonzero column 
and the $p$th entry $a_{p}b_{q}$ of the $q$th column be the upper most nonzero entry; 
we use zero-based numbering. 
Then we can see 
\[
a_{0}b_{q}=a_{1}b_{q}=\cdots=a_{p-1}b_{q}=0.
\]
Since $b_{q}\ne0$, 
$a_{0}=a_{1}=\cdots=a_{p-1}=0$, 
and so first $p$ rows of $B$ are all zero. 
Consider the diagonal passing the $(p,q)$ entry. 
Without loss of generality, let $p<q$, 
then we obtain 
\[
\alpha_{0,q-p}=\cdots=\alpha_{p,q}=0. 
\]
However, $0\ne a_{p}b_{q}= \alpha_{p-1,q-1}+\alpha_{p,q}=0$, 
this is a contradiction. 
Thus, $T^{\perp}$ has no rank-one matrices, 
and hence $\cC_0^{(1)}(T)=0$. 

We now show that $\cC_0^{(1)}(\Complex{I_2}\otimes{T})\ge\log{m}$. 
Let 
\[
\ket{\psi_i}=\ket{0}\ket{i}+\ket{1}\ket{i+1}\in\Complex^2\otimes\Complex^{m+1}, 
\]
where $0\le{i}\le{m-1}$. 
Then for any $R\in{T}$, 
\[
\bra{\psi_i}(I_2\otimes{R})\ket{\psi_j}=\bra{i}R\ket{j}+\bra{i+1}R\ket{j+1}=0
\]
Hence, $\cC_0^{(1)}({\Complex}I_2\otimes{T})\ge\log{m}$, 
\end{proof}

\begin{Rem}
When $m=3$ in Theorem~\ref{Thm:Act_eg}, 
we see that 
the one-shot zero-error classical capacity can be activated 
on $\Complex^{2}\otimes\Complex^{4}$. 
This result shows a lower dimensional case 
than the example in Ref.~\cite{Duan09} 
in which the input system is $\Complex^{2}\otimes\Complex^{6}$. 
Moreover, 
this example has the smallest input dimensions 
to be activated so far. 
\end{Rem}

Next, we show that 
the activation in theorem~\ref{Thm:Act_eg} also holds 
in the asymptotic setting. 
To do this, we need the following lemma 
based on Ref.~\cite{Duan_private13}.
\begin{Lem}~\label{Lem:having_rank-1} 
Let $S\le\Complex^{m_{1}\times{n_{1}}}$ and $T\le\Complex^{m_{2}\times{n_{2}}}$ be subspaces. 
Then $(S\otimes{T})^{\perp}$ has a rank-one matrix if and only if 
there exist nonzero matrices $A$ and $B$ such that $S \perp A\overline{T}B$. 
\end{Lem}
\begin{proof}
Suppose that $(S\otimes{T})^{\perp}$ has a rank-one matrix. 
Then there exist nonzero vectors $\ket{\psi}$ and $\ket{\phi}$ 
\begin{eqnarray*}
\ket{\psi}&=&\sum_{i=0}^{m_{1}-1}\sum_{j=0}^{m_{2}-1}a_{ij}\ket{i}\ket{j} \\
\ket{\phi}&=&\sum_{k=0}^{n_{1}-1}\sum_{l=0}^{n_{2}-1}b_{kl}\ket{k}\ket{l} 
\end{eqnarray*}
such that $\ket{\psi}\bra{\phi}\in(S\otimes{T})^{\perp}$. 
Then we can obtain 
for any $P\in{S}$ and $Q\in{T}$, 
\begin{equation}\label{eq:anOrthoElt}
\sum_{i,j,k,l}a_{ij}b_{kl}^{*}\bra{k}P^{\dag}\ket{i}\bra{j}\overline{Q}\ket{l}=0.
\end{equation}
Define two nonzero matrices 
\begin{eqnarray*}
A&=&\sum_{i=0}^{m_{1}-1}\sum_{j=0}^{m_{2}-1}a_{ij}\ket{i}\bra{j} \\
B&=&\sum_{k=0}^{n_{1}-1}\sum_{l=0}^{n_{2}-1}b_{kl}^{*}\ket{l}\bra{k}.
\end{eqnarray*}
Then by Eq.~(\ref{eq:anOrthoElt}), we obtain 
\begin{equation}
\Tr[P^{\dag}A\overline{Q}B]=\sum_{i,j,k,l}a_{ij}b_{kl}^{*}
\bra{k}P^{\dag}\ket{i}\bra{j}\overline{Q}\ket{l}=0
\end{equation}
for any $P\in{S}$ and $Q\in{T}$.
Similarly, we can readily see the converse.
\end{proof}

\begin{Thm}\label{Thm:Act_eg_asymp}
The noncommutative graphs $T$ in the proof of Theorem~\ref{Thm:Act_eg} cannot cause the superactivation. 
In particular, $\cC_0(T)=0$ 
and $\cC_0(\Complex{I_2}\otimes{T})\ge\log{m}>1=\cC_0(\Complex{I_2})+\cC_0(T)$.
\end{Thm}
\begin{proof} 
Let 
\[
T=\mathrm{span}\{B_{ij} : 0\le{i}\ne{j}\le{m-1}\}^{\perp},
\] 
where 
\[
B_{ij}=\ket{i}\bra{j}+\ket{i+1}\bra{j+1}\in\Complex^{(m+1)\times(m+1)}.
\]
Suppose that $\cC_0^{(1)}(S\otimes{T})>0$ 
for some noncommutative graph $S$ with $\cC_0^{(1)}(S)=0$. 
By Lemma~\ref{Lem:having_rank-1}, 
there exist nonzero matrices $A$ and $B$ such that 
$A\overline{T}B\subseteq{S^{\perp}}$. 
Since $\cC_0^{(1)}(S)=0$, 
\begin{equation}\label{eq:notRank=1}
\mathrm{rank}(A\overline{Q}B)\ne1 
\end{equation}
for any $Q\in T$. 

Let $Q\equiv\sum_{u,v=0}^{m}a_{uv}\ket{u}\bra{v}$ be any element in $T$. 
Then 
\[
0=\Tr{B_{ij}^{\dag}Q}=a_{i,j}+a_{i+1,j+1} 
\]
for $0\le{i}\ne{j}\le{m-1}$. 
From this, we can see that the followings belong to $T$: 
\begin{eqnarray}
\sum_{k=0}^{m-j}(-1)^{k}\ket{k}\bra{j+k}&,& j=1,\dots,m, \label{eq:diagonal_1} \\
\ket{i}\bra{i}&,& i=0,\dots,m, \label{eq:diagonal_2} \\
\sum_{k=0}^{m-j}(-1)^{k}\ket{j+k}\bra{k}&,& j=1,\dots,m. \label{eq:diagonal_3}
\end{eqnarray}
Putting matrices 
in Eqs.~(\ref{eq:diagonal_1}), (\ref{eq:diagonal_2}), and (\ref{eq:diagonal_3}) 
into Eq.~(\ref{eq:notRank=1}), 
we can see that 
there is $0\le{c}\le{m}$ such that 
$A\ket{i}=0$ for $i=0,\dots,c$ 
and $\bra{j}B=0$ for $j=c+1,\dots,m$. 
Similarly, there is $0\le{d}\le{m}$ such that 
$A\ket{i}=0$ for $i=d+1,\dots,m$ 
and $\bra{j}B=0$ for $j=0,\dots,d$. 
Then $A=0$ if $c\ge{d}$, and $B=0$ if $c\le{d}$. 
This is a contradiction since $A$ and $B$ are nonzero matrices. 
Therefore,  $\cC_0^{(1)}(S\otimes T)=0$ 
for any noncommutative graph $S$ with $\cC_0^{(1)}(S)=0$. 
\end{proof}

\begin{Rem}
When the zero-error classical capacity can be activated, 
we can raise the following question: 
how much can it be activated? 
In other words, 
how large can $\cC_0^{(1)}(S\otimes{T})-\cC_0^{(1)}(S)$ be 
for any noncommutative graph $T$ such that $\cC_0^{(1)}(T)=0$? 
In Theorems~\ref{Thm:Act_eg} and~\ref{Thm:Act_eg_asymp} 
as well as examples in Ref.~\cite{Duan09}, 
the capacity of the combined channel can be unbounded above, 
and so it may need to be regularized 
by the dimensions of systems. 
Then the (regularized) largest value could measure 
the ultimate ability to activate another useless quantum channel. 
The above-mentioned question is related with 
the concept of \emph{potential capacity} in Ref.~\cite{WY16}. 
\end{Rem}

%%%%%%%%%%%%%%%%%%%%%%%%%%%%%%%%%%%%%%%%%%%%
%%                                               Conclusions                                                            %%
%%%%%%%%%%%%%%%%%%%%%%%%%%%%%%%%%%%%%%%%%%%%
\section{Conclusions}\label{sec:conclusion} 
% summary %
We have considered 
when the activation of the zero-error classical capacity happens 
in low-dimensional input systems. 
First, we have shown that 
when one of two quantum channels is on a qubit system, 
the zero-error classical capacity of the combined channel can be activated 
only if the quantum channel on a qubit system is noiseless; 
that is, 
only a noiseless qubit channel can generate the activation. 
Moreover, we have shown that 
the zero-error classical capacity of two quantum channels on qubit systems 
cannot be activated. 
Finally, we have presented a class of examples 
showing the activation of the zero-error classical capacity 
in low-dimensional input systems. 
In particular, we have constructed 
an example having the smallest input dimensions so far.

\acknowledgments
We are thankful Runyao Duan and Soojoon Lee for helpful comments.
This research was supported by Basic Science Research Program 
through the National Research Foundation of Korea (NRF) 
funded by the Ministry of Education (2016R1A6A3A11936376), 
%This work was supported 
and by the ICT Research and Development program of MSIP/IITP 
[R0190-15-2030, 
Reliable crypto-system standards and core technology development for secure quantum key distribution network]. 

\bibliography{Act}

\end{document}